\documentclass{llncs}
\usepackage{llncsdoc}

\usepackage{graphicx}
\usepackage[caption=false]{subfig}
\usepackage{amsmath}
\usepackage{tabulary}

\newcommand{\DSS}{\texttt{DSS}}
\newcommand{\SSS}{\texttt{SSS}}
\newcommand{\RE}{\texttt{RE}}

\title{Drawing Big Graphs using Spectral Sparsification}
\author{Peter Eades, Quan Nguyen, Seok-Hee Hong}
\institute{The School of Information Technologies, University of Sydney Australia \\ \email{\{peter.eades,quan.nguyen,seokhee.hong\}@sydney.edu.au}}

\begin{document}

\maketitle

\begin{abstract}
Spectral sparsification is a general technique developed by Spielman \textit{et al.} to reduce the number of edges in a graph while retaining its structural properties. We investigate the use of spectral sparsification to produce good visual representations of big graphs. We evaluate spectral sparsification approaches on real-world and synthetic graphs. We show that spectral sparsifiers are more effective than random edge sampling. Our results lead to guidelines for using spectral sparsification in big graph visualization.
\end{abstract}

\section{Introduction}

The problem of drawing very large graphs is challenging and has motivated a large body of research
(see~\cite{WICS:WICS1343} for a survey).
As the number of vertices and edges becomes larger,
layout algorithms become less effective. Further, runtime is increased both at the layout stage and at the rendering stage.
Recent work (for example \cite{proxy2017}) approaches the problem by replacing the original graph with a ``proxy graph''. The proxy graph is typically much smaller than the original graph, and thus layout and rendering is easier.
The challenge for the proxy graph approach is to ensure that the proxy graph is a good representation of the original graph; for visualization, we want the drawing of the proxy graph to be \emph{faithful}~\cite{NguyenEH12a} to the original graph.

In this paper we examine a specific proxy graph approach using \emph{spectral sparsification} as introduced by Spielman \textit{et al.}~\cite{DBLP:journals/cacm/BatsonSST13}:
roughly speaking, the \emph{spectrum} (that is, the eigenvalues of the Laplacian; see  \cite{DBLP:books/daglib/0037866}) of the proxy graph approximates the spectrum of the original graph.
Since the spectrum is closely related to graph-theoretic properties that are significant for graph drawing,
this kind of proxy seems to promise faithful drawings.

We report results of an empirically investigation of the application of spectral sparsification to graph drawing.
Specifically, we consider two closely related spectral sparsification techniques, one deterministic and one stochastic.
We consider the quality of drawings so produced, using real-world and synthetic data sets.
Quality is evaluated using the shape-based proxy graph metrics~\cite{proxy2017}. The results of spectral sparsification are compared to sparsifications obtained by simple random edge sampling.
Our investigation confirms the promise of spectral sparsification,
and shows that (overall) it is better than simple random edge sampling.

Section~\ref{sec:background} recalls the proxy graph approach, and shape-based quality metrics for large graph drawing.
Section~\ref{sec:spectrum} describes the application of spectral sparsification to graph visualization. Section~\ref{sec:exp} presents our experiments with spectral sparsification. The results of these experiments are presented and discussed in Section~\ref{sec:qualityResults}.
Section~\ref{sec:conclusion} concludes.

\section{Background \label{sec:background}}

\subsubsection{Proxy graphs and sparsification.}
\begin{figure}
  \centering
  \includegraphics[width=1\columnwidth]{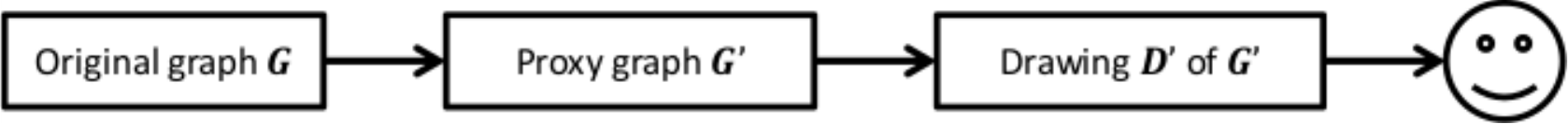}
  \caption{
  In the \emph{proxy graph approach}, the user sees a ``proxy'' of the original graph.
  }
  \label{fi:proxyPipeline}
\end{figure}

The proxy graph approach is described in Fig~\ref{fi:proxyPipeline}: for a given input graph $G$, a proxy graph $G'$ and a drawing $D'$ of $G'$ are computed. The proxy graph represents $G$ but is simpler and/or smaller than $G$ in some sense. The user sees the drawing $D'$ of $G'$, and does not see a drawing of the original graph $G$. However, if $G'$ is a ``good'' representation of $G$, then $D'$ is an adequate visualization of $G$ in that the user can see all the structure of $G$ in the drawing $D'$.

If $G'$ is a subgraph of $G$, and the edge density of $G'$ is smaller than the edge density of $G$, then we say that $G'$ is a \emph{sparsification} of $G$. Sparsification is the most common kind of proxy.

Sparsification has been extensively investigated in Graph Mining~\cite{gjoka2010walking,leskovec2006sampling,morstatter2013sample} (see survey~\cite{journals/corr/HuL13}). Typically, sparsification is achieved by some kind of stochastic sampling.
The most basic sparsification method is \emph{random edge sampling} \emph{(\RE)}:
each edge is chosen independently with probability $p$~\cite{rafiei2005effectively}.
This and many other simple stochastic strategies
have been empirically investigated in the context of visualization of large graphs~\cite{proxy2017,Wu2017sampling}.
In this paper we apply a more sophisticated graph sparsification approach to visualization: the \emph{spectral} sparsification work of Spielman \textit{et al.}~\cite{DBLP:journals/cacm/BatsonSST13,DBLP:journals/corr/abs-0803-0929,DBLP:journals/siamcomp/SpielmanT11}.

\subsubsection{Shape-based quality metrics. \label{sec:metrics}}
Traditional graph drawing metrics such as \emph{edge bends}, \emph{edge crossings}, and \emph{angular resolution} are based on the \emph{readability} of graphs; these metrics are good for small scale visualisation but become meaningless beyond a few hundred nodes~\cite{DBLP:conf/apvis/NguyenEH13}. For large graphs, \emph{faithfulness metrics} are more important: informally, a drawing $D$ of a graph $G$ is \emph{faithful} insofar as $D$ determines $G$, that is, insofar as the mapping $G \rightarrow D$ is invertible.

Here we use \emph{shape-based faithfulness metrics}~\cite{DBLP:journals/jgaa/EadesH0K17}. The aim of these metrics is to measure how well the ``shape'' of the drawing represents the graph.
\begin{figure*}\centering
	\captionsetup[subfloat]{farskip=2pt,captionskip=2pt}
	\setlength{\tabcolsep}{6pt}
	\begin{tabular}{|c|c|c|}
		\hline
		\subfloat[]{
			\includegraphics[width=0.25\linewidth]{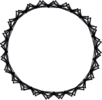}} &
		\subfloat[]{
			\includegraphics[width=0.25\linewidth]{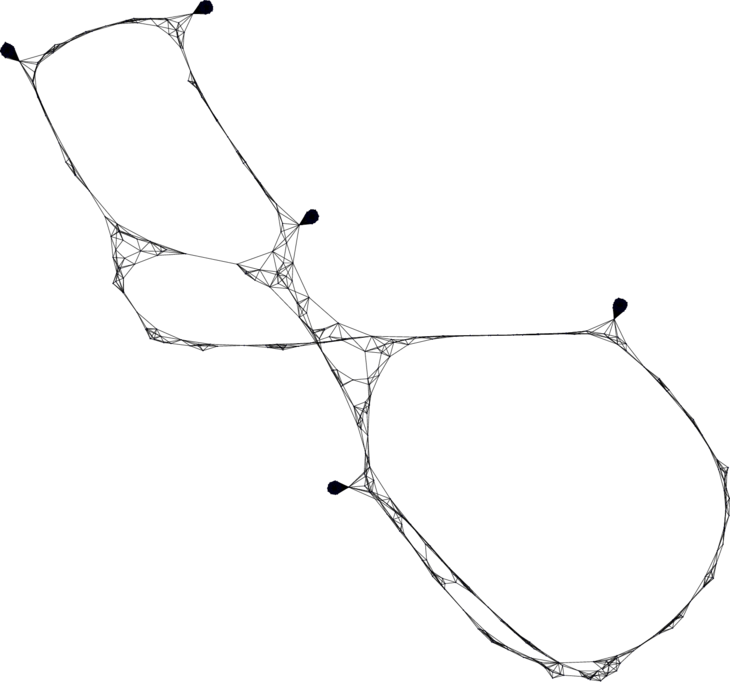}}	&		
		\subfloat[]{
			\includegraphics[width=0.25\linewidth]{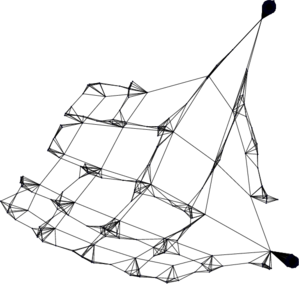}}
		\\\hline
	\end{tabular}
	\caption{The graphs \emph{can\_144}, \emph{cN1031M22638}, and \emph{gN733M62509}, drawn using FM$^3$. Note that each has a distinctive shape.\label{fig:originals}}	
\end{figure*}
For large graphs, such as in Fig.~\ref{fig:originals}, the shape of the drawing is more significant than the number of edge bends and edge crossings.
To make this notion more precise, we use ``shape graphs''. Given a set $P$ of $n$ points, a \emph{shape graph} $S(P)$ is a graph with vertex set $P$ such that the edge of $S(P)$ define the ``shape'' of $P$ in some sense. Examples of shape graphs are the \emph{Euclidean minimum spanning tree (EMST)}, the \emph{relative neighbourhood graph (RNG)}, and the \emph{Gabriel graph (GG)}~\cite{DBLP:journals/pr/Toussaint80}.

Suppose that $G = (V,E)$ is a graph and $P$ is a set of points in the plane,
and each vertex $u \in V$ is associated with a point $p(u) \in P$.
Denote the set of neighbours of $u$ in $G$ by $N_G(u)$, and the set of neighbours of $p(u)$ in the shape graph $S(P)$ by $N_{S(P)}(p(u))$.
We say that
\begin{equation*}
Jaccard(S(P),G) =
\frac{1}{|V|}
\sum_{u \in V} \frac{ | N_G(u) \cap N_{S(P)}(p(u)) |}{| N_G(u) \cup N_{S(P)}(p(u)) | }
\end{equation*}
is the \emph{Jaccard similarity} between the shape graph $S(P)$ and $G$.
If $D$ is a drawing of $G$ then the \emph{(shape-based) quality} of $D$ is $Q(D,G) = Jaccard(P,G)$, where $P$ is the set of vertex locations in the drawing $D$. Similarly, if $D'$ is a drawing of a sparsification $G'$ of $G$, then the \emph{(shape-based) (proxy) quality} of $D'$ is $Q(D',G) = Jaccard(P',G)$, where $P'$ is the set of vertex locations in the drawing $D'$. 
Note that if $u$ does not occur in $D'$, we consider that $N'(u)=\emptyset$.
For more details, see~\cite{EadesJGAA2017}.

\section{The spectral sparsification approach to large graph drawing \label{sec:spectrum}}

First we describe some of the terminology and concepts of spectral graph theory. More details are in standard texts; for example, \cite{FanChung,DBLP:books/daglib/0037866}\footnote{Beware: much of the terminology in spectral graph theory is not standardised.}.
The \emph{adjacency matrix} of an $n$-vertex graph $G=(V,E)$ is the $n \times n$ matrix $A$, indexed by $V$, such that $A_{uv} = 1$ if $(u,v) \in E$ and $A_{uv} = 0$ otherwise. The \emph{degree matrix} $D$ of $G$ is the diagonal matrix with where $D_{uu}$ is the degree of vertex $u$. The \emph{Laplacian} of $G$ is $L = D - A$.
The \emph{spectrum} of $G$ is the list
$\lambda_1 , \lambda_2, \ldots , \lambda_n$ of eigenvalues of $L$. It can be shown that $L$ has real nonnegative eigenvalues~\cite{FanChung}, and we assume that $\lambda_1 \leq \lambda_2 \leq \ldots \leq \lambda_n$; straightforward computation shows that $\lambda_1 = 0$.

The spectrum of a graph is closely related to many structural properties of the graph:
\begin{description}

 \item [Connectivity:]
 The number of connected components of $G$ is largest value of $i$ for which $\lambda_i$ = 0~\cite{FanChung}. Roughly speaking, a larger value of $\lambda_2$ indicates a more highly connected graph and is related to the diameter of the graph. If $\lambda_2 > 0$ then it is called the \emph{algebraic connectivity} of $G$~\cite{GrossAndYellen}.

 \item [Clusters:] \emph{Spectral clustering} involves the projection of the graph using its smallest eigenvalues. Spectral clustering solves a relaxation of the \emph{ratio cut} problem, that is, the problem of dividing a graph into clusters to minimise the ratio between the number of inter-cluster edges and the cluster size~\cite{vonLuxburg}. Informally, the ratio cut problem seeks to find clusters of similar size so that the coupling between clusters is minimised.
  \item [Stress:]
  The spectrum solves a kind of constrained stress problem for the graph. More specifically, the Courant - Fischer theorem (see~\cite{FanChung}) implies that
       \begin{equation}
       \label{eq:stress}
       \lambda_i = \min_{x \in X_i} \sum_{(u,v) \in E} (x_u - x_v)^2,
       \end{equation}
      where $X_i$ is the set of unit vectors orthogonal to the first ($i-1$) eigenvectors.
      The minimum is achieved when $x$ is an eigenvalue corresponding to $\lambda_i$. Note that the right hand side of equation~(\ref{eq:stress}) is a kind of stress function.
  \item [Commute distance:]
  The average time that a random walker takes to travel from vertex $u$ to vertex $v$ and return is the \emph{commute distance} between $u$ and $v$. Eigenvalues are related to random walks in the graph, and thus to commute distances (see \cite{Lovasz}).
\end{description}
Spielman and Teng~\cite{DBLP:journals/siamcomp/SpielmanT11}, following Benczur and Karger~\cite{DBLP:conf/stoc/BenczurK96}, first introduced the concept of ``spectral approximation''.
Suppose that $G$ is an $n$-vertex graph with Laplacian $L$, and $G'$ is an $n$-vertex subgraph of $G$ with Laplacian $L'$.
If there is an $\epsilon > 0$ such that for every $x \in R^n$,
\begin{equation}
\label{eq:approximation}
(1 - \epsilon)\frac{x^T L 'x}{x^T x} \leq \frac{x^T Lx}{x^T x} \leq (1 + \epsilon) \frac{x^T L' x}{x^T x},
\end{equation}
then $G'$ is an \emph{$\epsilon$-spectral approximation} of $G$.
Using the Courant-Fischer Theorem~\cite{FanChung} with (\ref{eq:approximation}), one can show that if $G'$ is an \emph{$\epsilon$-spectral approximation} of $G$ then the eigenvalues and eigenvectors of $G'$ are close to those of $G$.
The importance of this is that spectral approximation preserves the structural properties listed above.

Spielman and Teng first showed that every $n$-vertex graph has a spectral approximation with $O(n \log n)$ edges~\cite{DBLP:journals/siamcomp/SpielmanT11}.
The following theorem is one such result:
\begin{theorem}
\label{th:SpielmanTeng}\cite{DBLP:journals/siamcomp/SpielmanT11}
Suppose that  $G$ is an $n$-vertex graph and $\frac{1}{\sqrt{n}} \leq \epsilon \leq 1$.
Then with probability at least $\frac{1}{2}$, there is an $\epsilon$-spectral approximation $G'$ of $G$
with $O(\frac{1}{\epsilon} n \log n)$ edges.
\end{theorem}
Further research of Spielman \textit{et al.} refines and improves spectral sparsification methods (see~\cite{DBLP:journals/cacm/BatsonSST13}). These results have potential for resolving scale issues in graph visualisation by reducing the size of the graph while retaining its (spectral) structure.
However, the \emph{practical impact} of these results for graph visualization is not clear, because of large constants involved. 

The proof of Theorem~\ref{th:SpielmanTeng} is essentially a stochastic sampling method, using the concept of ``effective resistance''. Suppose that we regard a graph $G=(V,E)$ as an electrical network where each edge is a 1-$\Omega$ resistor, and a current is applied. The voltage drop over an edge $(u,v)$ is the \emph{effective resistance} $r_{uv}$ of $(u,v)$. Effective resistance in a graph is closely related to commute distance, and can be computed simply from the Moore-Penrose inverse~\cite{Ben-Israel} of the Laplacian.
If $L^{\dag}$ is the Moore-Penrose inverse of $L$ and $(u,v) \in E$, then  $r_{uv} = L^{\dag}_{uu}$ + $L^{\dag}_{vv}$ - 2 $L^{\dag}_{uv}$.

We next describe two graph drawing algorithms, both variants of algorithms of Spielman \textit{et al.}~\cite{DBLP:journals/cacm/BatsonSST13}.
Each takes a graph $G$ and an integer $m'$, and computes a sparsification $G'$ with $m'$ edges, then draws $G'$.
\begin{description}
\item[\SSS] (\underline{S}tochastic \underline{S}pectral \underline{S}parsification) randomly selects edges with probability proportional to their resistance value. Let $E'$ be the edge set from $m'$ random selections.
Let $G'$ be the subgraph of $G$ induced by $E'$; draw $G'$.
\item[\DSS] (\underline{D}eterministic \underline{S}pectral \underline{S}parsification). Let $E'$ consist of the $m'$ of largest effective resistance. Let $G'$ be the subgraph of $G$ induced by $E'$; draw $G'$.
\end{description}
In both \DSS~and \SSS, the sparsified graph can be drawn with any large-graph layout algorithm.

\section{The experiments \label{sec:exp}}

The driving hypothesis for this paper is that for large graphs,
spectral sparsification gives good proxy graphs for visualization.
To be more precise, we define the \emph{relative density} of the sparsification
$G'$ for a graph $G$ to be $\frac{m'}{m}$, where $G$ has $m$ edges and $G'$ has $m'$ edges.
Note that a proxy with higher relative density should be a better approximation to the original graph;
thus we expect that drawings of the proxy with higher relative density should have better quality.

Since spectral sparsification (approximately) preserves the eigenvalues,
we hypothesize that both \SSS~and \DSS~are better than \RE. Further, we expect that
the difference becomes smaller when the relative density is larger.
To state this precisely, let $D'_{\SSS}$ (respectively $D'_{\RE}$) denote the drawing obtained by \SSS~(respectively \RE). We say that $\frac{Q(D'_{\SSS},G)}{Q(D'_{\RE}, G)}$ is the \emph{quality ratio} of \SSS; similarly define the quality ratio of \DSS.
We expect that the quality ratio of both \SSS~and \DSS~is greater than 1.
Further, we expect that the quality ratio for both algorithms tends to 1
as relative density tends to 1.

We implemented \DSS, \SSS~and \RE~in Java, on top of the OpenIMAJ toolkit~\cite{DBLP:conf/mm/HareSD11}.
In particular, we used OpenIMAJ to compute the Moore-Penrose inverse.
The experiments were performed on a Dell XPS 13 laptop, with an i7 Processor, 16GB memory and 512GB SSD. The laptop was running Ubuntu 16.04 with 20GB swap memory. The computation of the Moore-Penrose inverse used Java 8, with a specified 16GB heap. We used multiple threads to speed up the resistance computation.

We used three data sets.
The first set of graphs is taken from ``defacto-benchmark'' graphs,
including the Hachul library, Walshaw's Graph Partitioning Archive, the sparse matrices collection~\cite{Davis:2011} and the network repository~\cite{nr-aaai15}. These include two types of graphs that have been extensively studied in graph drawing research: grid-like graphs and scale-free graphs.
The second set is the GION data set~\cite{DBLP:conf/gd/MarnerSTKEH14};
this consists of RNA sequence
graphs that are used for the analysis of repetitive sequences in sequencing data;
these graphs have been used in previous experiments.
They are locally dense and globally sparse, and generally have distinctive shapes.
The third set consists of randomly generated graphs that contain interesting structures that are difficult to model with sparsification. Specifically, we generated a number of ``black-hole graphs'', each of which consists of one or more large and dense parts (so-called ``black holes''), and these parts connect with the rest of the graph by relatively few edges. These relatively few  edges \emph{outside} the ``black holes'' determine the structure of the graph. Such graphs are difficult to sparsify because sampling strategies tend to take edges from the dense ``black holes'' and miss the structurally important edges. Figs.~\ref{fig:originals}(b) and (c) are black-hole graphs.
Details of the graphs that we used are in Table~\ref{table:data}.
\begin{table}[t]
	\centering
	\caption{\label{table:data}Data sets}
	\subfloat[Benchmark graphs]{
		\begin{tabular}{|l|c|c|c|c|}
		\hline
		graph &	$|V|$	& $|E|$ & type \\
		\hline
		can\_144 & 144 &	576 & grid \\
		G\_15 	& 1785	& 20459 & scalef \\
		G\_2 	& 4970	& 7400 & grid \\
		G\_3	& 2851	& 15093 & grid \\	
		G\_4	& 2075	& 4769 & scalef \\	
		mm\_0	& 3296	& 6432 & grid \\	
		nasa1824	& 1824	& 18692 & grid  \\	
		facebook01 & 4039	& 88234 & scalef \\
		oflights & 2939	& 15677	& scalef \\
		soc\_h & 2426	& 16630 & scalef	\\
		yeastppi & 2361	& 7182 & scalef \\
		\hline
		\end{tabular}	
	}	
	\hfill
	\subfloat[GION graphs]{
		\begin{tabular}{|l|c|c|}
			\hline
			graph & $|V|$	& $|E|$ \\
			\hline	
		graph\_1  & 5452 & 118404\\
		graph\_2  & 1159 & 6424\\
		graph\_3  & 7885 & 427406\\
		graph\_4  & 5953 & 186279\\
		graph\_5  & 1748 & 13957\\
		graph\_6  & 1785 & 20459\\
		graph\_7  & 3010 & 41757\\	
		graph\_8 & 4924 & 52502\\
		\hline
		\end{tabular}
	}
	\hfill
	\subfloat[Black-hole graphs]{
		\begin{tabular}{|l|c|c|}
			\hline
			graph & $|V|$	& $|E|$ \\
			\hline
			cN377M4790  & 377 & 4790\\
			cN823M14995  & 823 & 14995\\
			cN1031M226386  & 1031 & 22638\\
			gN285M2009 & 285 & 2009\\
			gN733M62509  & 733 & 62509\\
			gN1080M17636 & 1080 & 17636\\
			gN4784M38135 & 4784 & 38135\\
			\hline
		\end{tabular}
	}		
\end{table}

We sparsify these input graphs to a range of relative density values: from small (1\%, 2\%, 3\%, 4\%, 5\%, 10\%) to medium and large (15\%, 20\%, $\cdots$, 100\%), using \SSS, \DSS, and \RE.

For layout, we use the \emph{FM$^3$} algorithm~\cite{Hachul:2004}, as implemented in OGDF~\cite{MarkusChimani2012}. However, we also confirmed our results using FM$^3$ variants (see Section~\ref{sec:results}).

We measured quality of the resulting visualizations by proxy quality metrics $Q(D,G)$ described in Section~\ref{sec:metrics}.
For shape graphs, we used \emph{GG}, \emph{RNG}, and \emph{EMST}; the results for these three shape graphs are very similar, and here we report the results for \emph{GG}.

\section{Results from the experiments\label{sec:qualityResults}}


First we describe typical examples of the results of our experiments, using the graphs illustrated in Fig.~\ref{fig:originals}; these are a relatively small defacto-benchmark graph \emph{can\_144}, and two black-hole graphs \emph{cN1031M22638} and \emph{gN733M62509}.

\begin{figure*}[t!]\centering
	\captionsetup[subfloat]{farskip=2pt,captionskip=2pt}
	\setlength{\tabcolsep}{6pt}
	\begin{tabular}{|c|c|c|}
		\hline
		\subfloat[\RE~ 3\%]{
			\includegraphics[width=0.22\linewidth]{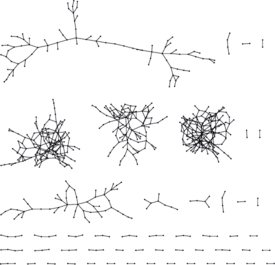}}		
		&
		\subfloat[\DSS~3\%]{
			\includegraphics[width=0.22\linewidth]{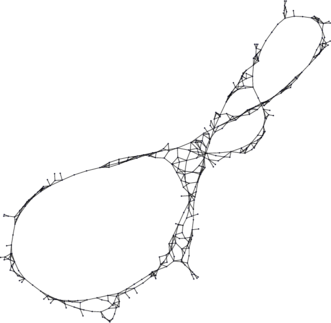}}
		&
		\subfloat[\SSS~3\%]{
			\includegraphics[width=0.22\linewidth]{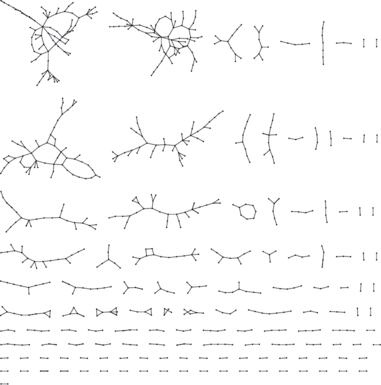}}
		\\\hline		
		\subfloat[\RE~ 15\%]{					
			\includegraphics[width=0.22\linewidth]{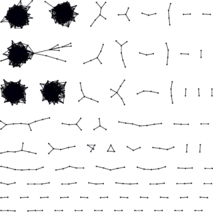}}
		&
		\subfloat[\DSS~15\%]{
			\includegraphics[width=0.22\linewidth]{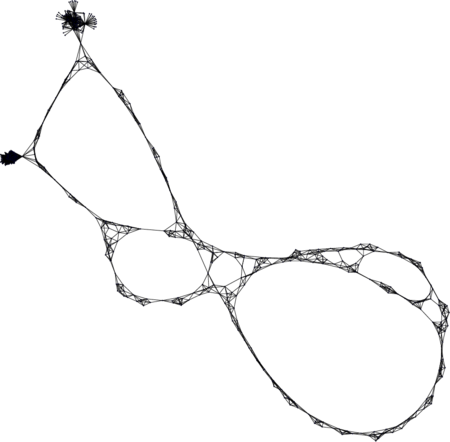}}
		&
		\subfloat[\SSS~15\%]{
			\includegraphics[width=0.22\linewidth]{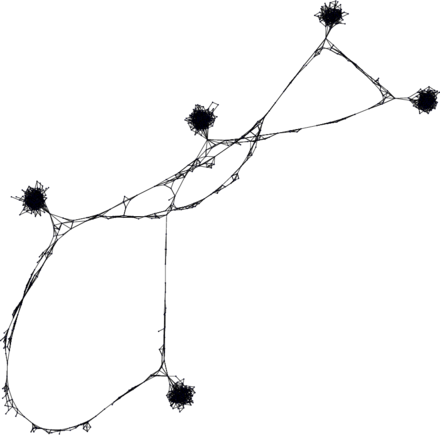}}
		\\\hline
	\end{tabular}	
	\caption{Sparsifications of the graph cN1031M22638 at relative densities 3\% and 15\%.\label{fig:cN1031M22638}}
\end{figure*}

\begin{figure*}[t!]\centering
	\captionsetup[subfloat]{farskip=2pt,captionskip=2pt}
	\setlength{\tabcolsep}{6pt}
	\begin{tabular}{|c|c|c|}
		\hline		
		\subfloat[\RE~ 1\%]{
			\includegraphics[width=0.22\linewidth]{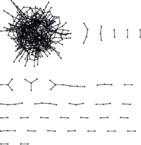}} &		
		\subfloat[\DSS~1\%]{
			\includegraphics[width=0.22\linewidth]{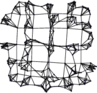}} &
		\subfloat[\SSS~1\%]{
			\includegraphics[width=0.22\linewidth]{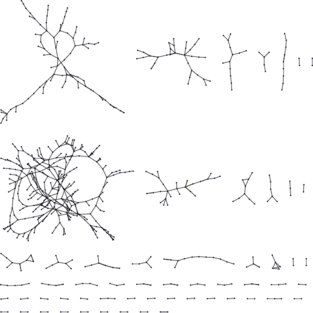}}
		\\\hline
		\subfloat[\RE~ 10\%]{					
			\includegraphics[width=0.22\linewidth]{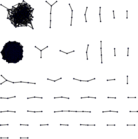}} &	
		\subfloat[\DSS~10\%]{
			\includegraphics[width=0.22\linewidth]{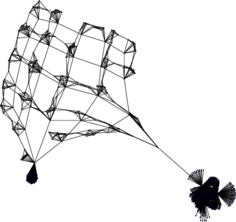}} &			
		\subfloat[\SSS~10\%]{
			\includegraphics[width=0.22\linewidth]{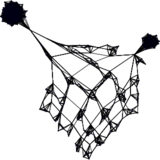}}
		\\\hline
	\end{tabular}	
	\caption{Sparsifications of the graph \emph{gN733M62509} at relative densities 1\% and 10\%.\label{fig:gN733M62509}}
\end{figure*}

Sparsifications of \emph{cN1031M22638} using \RE, \DSS, and \SSS~at relative densities of 3\% and 15\% are in Fig.~\ref{fig:cN1031M22638}.
At relative density of 3\%, both \RE~ and \SSS~give poor results; the drawings do not show the structure of the graph. However, \DSS~gives a good representation. At relative density 15\%, both \DSS~and \SSS~are good, while \RE~remains poor. A similar example, with relative densities of 1\% and 10\% for the black-hole graph \emph{gN733M62509}, is in Fig.~\ref{fig:gN733M62509}.


\begin{figure*}[t!]\centering
	\captionsetup[subfloat]{farskip=2pt,captionskip=2pt}
	\setlength{\tabcolsep}{6pt}
	\begin{tabular}{|c|c|c|}
		\hline
		\subfloat[\RE~ 40\%]{
			\includegraphics[width=0.22\linewidth]{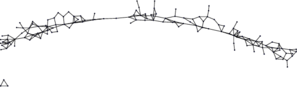}} &		
		\subfloat[\DSS~40\%]{
			\includegraphics[width=0.22\linewidth]{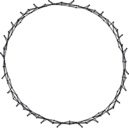}} &
		\subfloat[\SSS~40\%]{
			\includegraphics[width=0.22\linewidth]{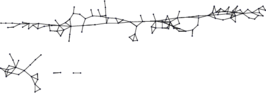}}
		\\\hline
		\subfloat[\RE~ 50\%]{
			\includegraphics[width=0.22\linewidth]{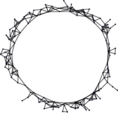}} &	
		\subfloat[\DSS~50\%]{
			\includegraphics[width=0.22\linewidth]{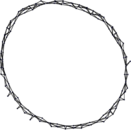}} &
		\subfloat[\SSS~50\%]{
			\includegraphics[width=0.22\linewidth]{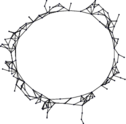}}		
		\\\hline
	\end{tabular}	
	\caption{Sparsifications of the graph \emph{can\_144} at relative densities 40\% and 50\%.\label{fig:can144}}
\end{figure*}

While the results for \emph{cN1031M22638} and \emph{gN733M62509} are typical, some results did not fit this mold. For \emph{can\_144}, see  Fig.~\ref{fig:can144}; here \RE~and \SSS~give poor representations, even at very high relative density (40\%). However, all three algorithms give good representations at relative density 50\%.

\subsection{Quality: results and observations}
\label{sec:results}

\begin{figure*}[t]\centering
	\captionsetup[subfloat]{farskip=2pt,captionskip=2pt}
	\subfloat[$Q_{\DSS}$ - Benchmark graphs\label{fig:RESA-real}]{
		\includegraphics[width=0.4\linewidth]{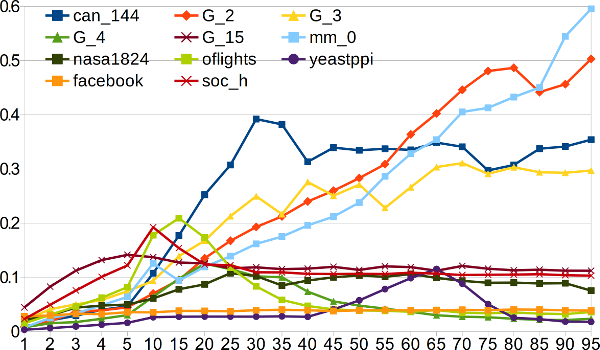}}
	\hspace{1cm}	
	\subfloat[$Q_{\SSS}$ - Benchmark graphs\label{fig:SSS-real}]{
		\includegraphics[width=0.4\linewidth]{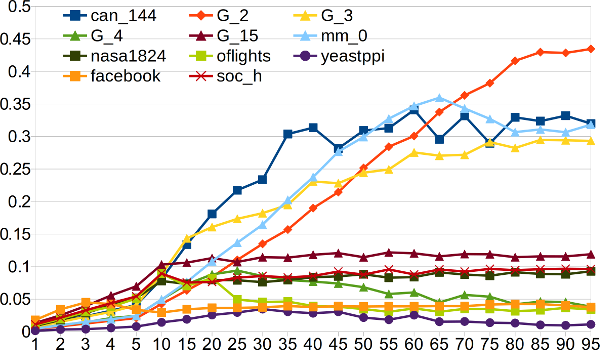}}
	\\
	\subfloat[$Q_{\DSS}$ - GION graphs \label{fig:RESA-GION}]{
		\includegraphics[width=0.4\linewidth]{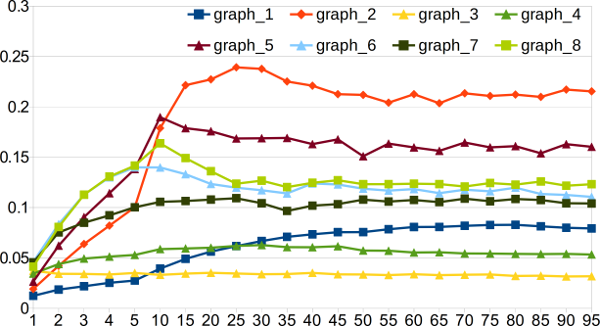}}
	\hspace{1cm}	
	\subfloat[$Q_{\SSS}$ - GION graphs \label{fig:SSS-GION}]{
		\includegraphics[width=0.4\linewidth]{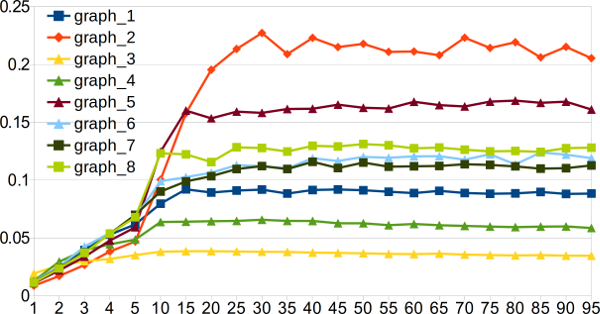}}
	\\
	\subfloat[$Q_{\DSS}$ - Black-hole graphs \label{fig:RESA-blackhole}]{
		\includegraphics[width=0.4\linewidth]{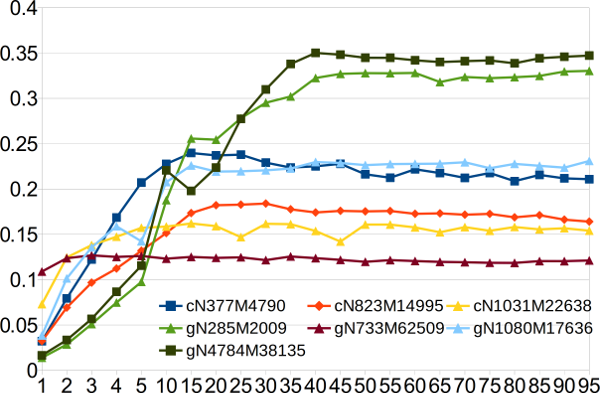}}
	\hspace{1cm}
	\subfloat[$Q_{\SSS}$ - Black-hole graphs \label{fig:SSS-blackhole}]{
		\includegraphics[width=0.4\linewidth]{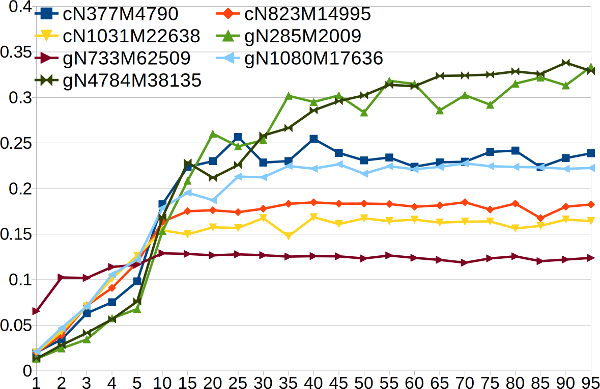}}
	\caption{Proxy quality metrics of output of \DSS, \SSS~and  \RE: (a)-(b) defacto-benchmark graphs, (c)-(d) GION graphs, (e)-(f) black-hole graphs. The metrics use FM$^3$ layout and GG shape graphs.\label{fig:metricsVSsiz}}
\end{figure*}

Fig.~\ref{fig:metricsVSsiz} shows the quality metrics for the three data sets for all three algorithms.
The $x$-axis shows relative densities from 1\% to 95\%; the $y$-axis shows quality measures of the proxies.

We make the following five observations from the results.

\begin{figure*}[t]\centering
	\captionsetup[subfloat]{farskip=2pt,captionskip=2pt}	
	\subfloat[$Q_{\DSS}/Q_{\RE}$ - Benchmark graphs]{\includegraphics[width=0.4\linewidth]{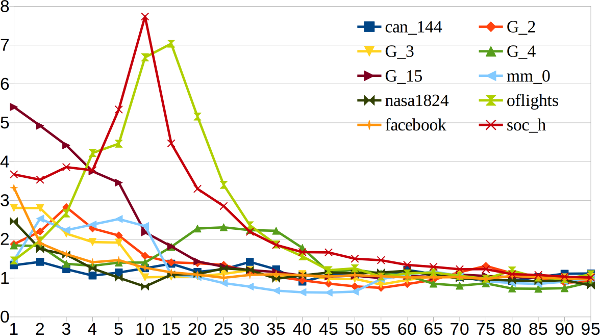}}
	\hspace{1cm}	
	\subfloat[$Q_{\SSS}/Q_{\RE}$ - Benchmark graphs]{
		\includegraphics[width=0.4\linewidth]{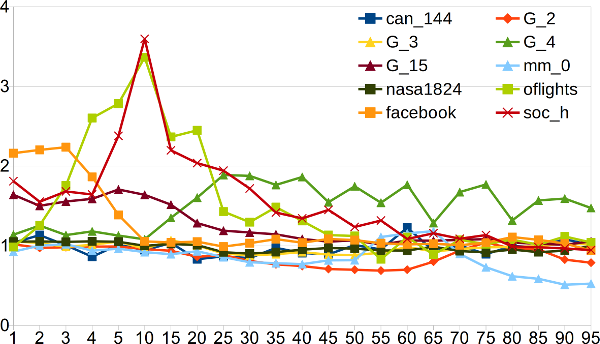}}
	\\
	\subfloat[$Q_{\DSS}/Q_{\RE}$ - GION graphs]{
		\includegraphics[width=0.4\linewidth]{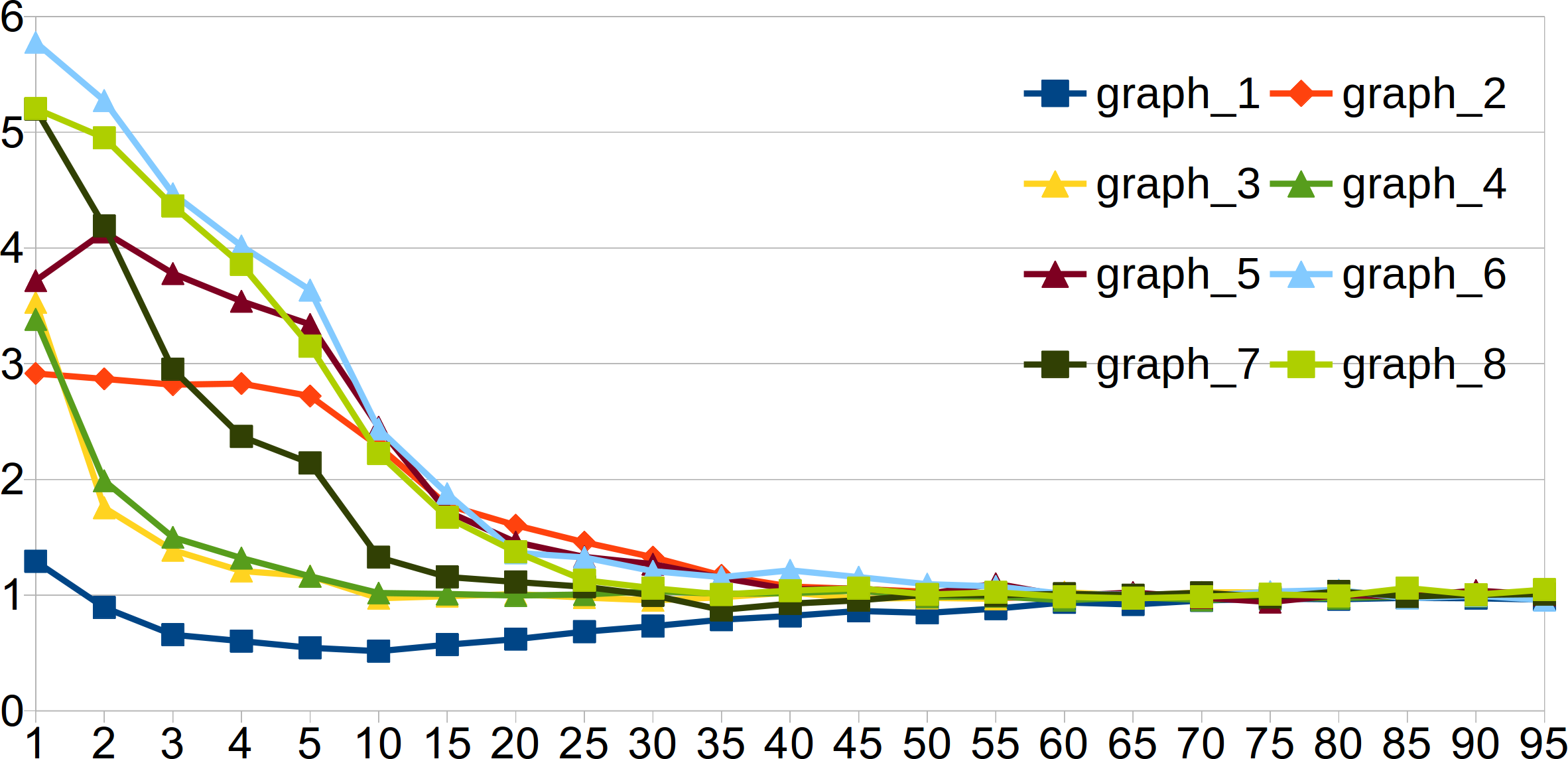}}
	\hspace{1cm}	
	\subfloat[$Q_{\SSS}/Q_{\RE}$ - GION graphs]{
		\includegraphics[width=0.4\linewidth]{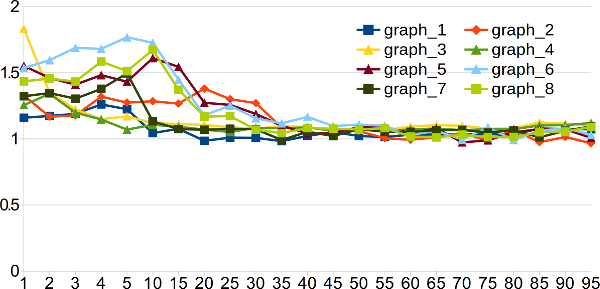}}
	\\
	\subfloat[$Q_{\DSS}/Q_{\RE}$ - Black-hole graphs]{
		\includegraphics[width=0.4\linewidth]{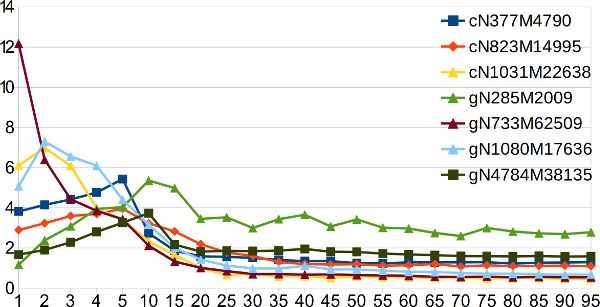}}
	\hspace{1cm}
	\subfloat[$Q_{\SSS}/Q_{\RE}$ - Black-hole graphs]{
		\includegraphics[width=0.4\linewidth]{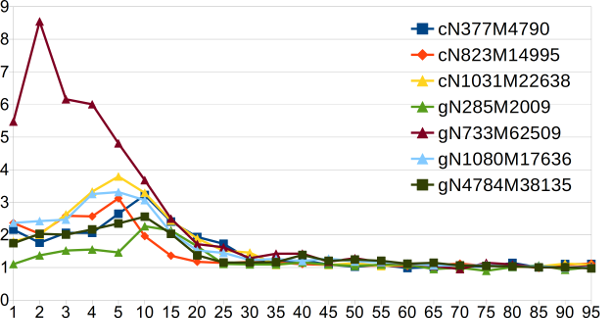}}
	\caption{\label{fig:improveReal}Proxy quality ratio for \DSS~ and \SSS, for (a)-(b) defacto-benchmark graphs, (c)-(d) GION graphs, (e)-(f) black hole graphs. The $y$-axis shows the quality ratio.}
\end{figure*}

\begin{enumerate}
\item
\textbf{Quality increases with relative density.}
In general, quality increases as relative density increases.
For many graphs there is a more interesting pattern: quality mostly increases up to a limit,
achieved at a relative density between 10\% and 30\%, and then stays steady.
Some of the defacto-benchmark graphs do not show this pattern:
they show close to linear improvement in quality with density all the way up to 95\%.


\item
\textbf{Spectral sparsification is better than random edge sampling.}

Fig.~\ref{fig:improveReal} depicts the quality ratio ($y$-axis) for \DSS~and \SSS~for each data set, over
relative density from 1\% to 95\%.
Note that the quality ratio is significant in most cases, especially at low relative density. For example, \DSS~metrics are around 200 times better than \RE, and sometimes much more (for the yeast dataset it is about 400).

For most of the graphs, the quality ratio decreases as the relative density increases. Quality ratio is best for relative density smaller than 10\%. When the relative density is more than 15\%,  \RE~may be slightly better than \DSS~for a few graphs, such as defacto-benchmark graphs $mm\_0$ graph (light blue), and $G\_2$ (red). Interestingly, $soc\_h$ and $oflights$ show a peak at around 10\% and 15\% before a drop for larger relative density.


\item
\textbf{Sparsification is better for grid-like graphs than for scale-free graphs.}
Fig.~\ref{fig:qualityAvgReal}(a) shows the quality change for \DSS, \SSS, and \RE with density, over the grid-like and scale-free defacto-benchmark graphs.
\begin{figure*}[htp]\centering
	\subfloat[Average quality metrics]{
		\includegraphics[width=0.45\linewidth]{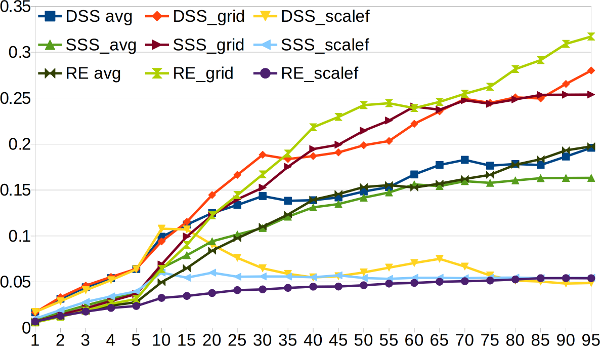}}
	\hfill
	\subfloat[Average quality ratio]{
		\includegraphics[width=0.4\linewidth]{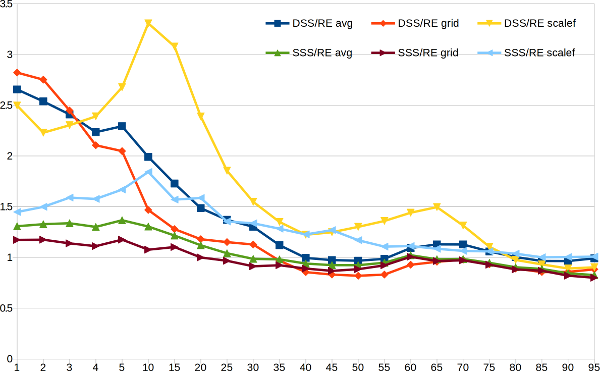}}
	\caption{\label{fig:qualityAvgReal} Comparison of proxy quality metrics of defacto-benchmark graphs: (1) Average quality measures, (2) Average of quality ratio.  The values are computed by graph types for scale-free graphs (\_scalef), grid-like (\_grid) graphs, and overall (\_avg). $Q_{\DSS}/Q_{\RE}$.}
\end{figure*}
Note that average values for \DSS~and \SSS~are better than the average value for \RE~when the relative density is less than 35\%. When relative density is greater than 40\%, there are fluctuations between \SSS~and \DSS. For grid-like graphs, the \DSS~and \SSS~proxies give better average proxy measures than \RE~proxies for relative density less than 20\%. For relative density greater than 35\%, \RE~proxies improve.
For scale-free graphs, \DSS~and \SSS~outperformed when relative density is under 80\%.

Fig.~\ref{fig:qualityAvgReal}(b) shows the ratio of the quality average between \DSS~over \RE~and \SSS~over \RE.
Overall, the quality ratios decline when relative density increase. The ratios are good from 1.2 to 3 times better for relative density up to 20\%. For both types of graphs, \DSS~gives best quality, then \SSS~comes second.


\item
\textbf{Deterministic spectral sparsification is better.}
We compared the average of quality metrics for \DSS, \SSS~and \RE~sparsification.
Fig.~\ref{fig:avgmetrics} shows the average quality values for the three data sets.
As expected, average values increase when the relative density increases.
Note that \DSS~gives the best average and \SSS~is the second best.

\begin{figure*}[t!]\centering	
	\subfloat[Benchmark graphs]{
		\includegraphics[width=0.32\linewidth]{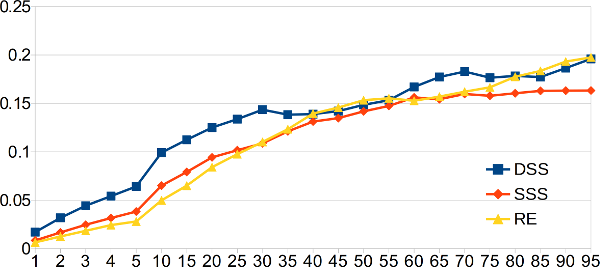}}
	\subfloat[GION graphs]{
		\includegraphics[width=0.32\linewidth]{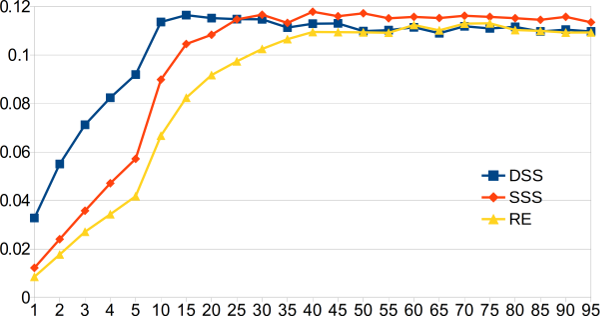}}
	\subfloat[Black-hole graphs]{
		\includegraphics[width=0.32\linewidth]{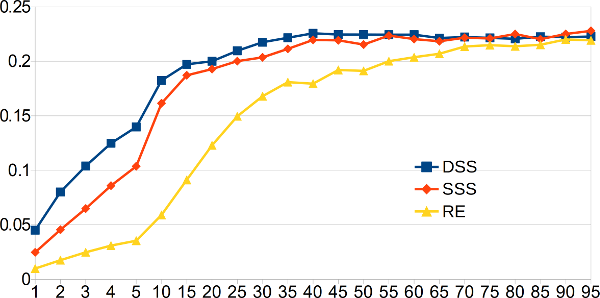}}
	\caption{Average quality metrics of \DSS, \SSS~and  \RE~over all data sets.\label{fig:avgmetrics}}
\end{figure*}

\begin{figure*}[t!]\centering	
	\subfloat[Benchmark graphs]{
		\includegraphics[width=0.32\linewidth]{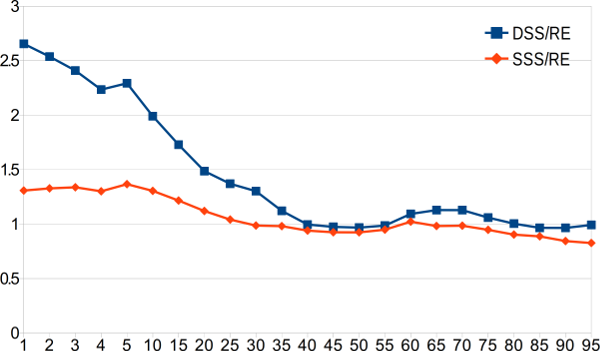}}
	\subfloat[GION graphs]{	
		\includegraphics[width=0.32\linewidth]{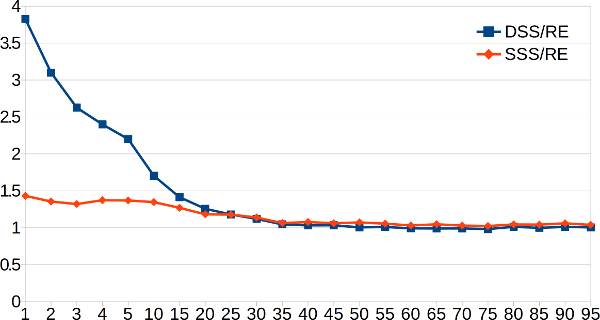}}
	\subfloat[Black-hole graphs]{
		\includegraphics[width=0.32\linewidth]{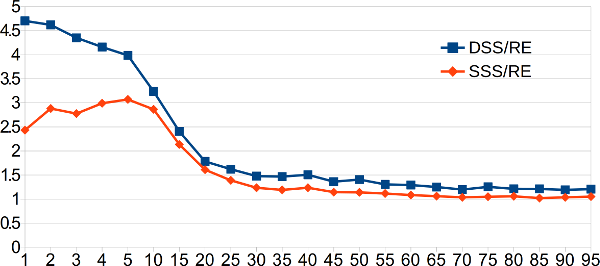}}
	\caption{Quality ratios of \DSS/\RE~and \SSS/\RE~over all data sets.\label{fig:avgratio}}
\end{figure*}

Fig.~\ref{fig:avgratio} shows the quality ratios  $Q_{\DSS}/Q_{\RE}$ and  $Q_{\SSS}/Q_{\RE}$ for all the data sets. Again, \DSS~gives an overall larger improvement over \RE~than \SSS. The improvement of \DSS~over \RE~is good when relative density is less than 35\%; \SSS~shows in improvement over \RE~as well, but it is not so dramatic. When relative density is beyond 35\%, the ratio becomes small (close to 1) or even becomes smaller than 1. Further note from
Fig.~\ref{fig:avgratio}(a)-(c) that \DSS~and \SSS~give better quality ratios for black-hole graphs than for GION graphs and defacto-benchmark graphs.


\item
\textbf{Quality results are consistent across different layout algorithms.}
The results reported above use FM$^3$ for layout. However, we found that results using other layout algorithms were very similar.
We measured the quality ratios using \emph{FM$^3$}, \emph{Fast}, \emph{Nice} and \emph{NoTwist} layouts from OGDF. For example,
Fig.~\ref{fig:qualityratioLayouts} shows the quality ratio of \DSS. As depicted from the graphs, the improvement of \DSS~over \RE~is consistent across different layout algorithms. The differences in the ratios is very small.

\begin{figure*}[t!]\centering	
	\subfloat[GION graphs]{
		\includegraphics[width=0.48\linewidth]{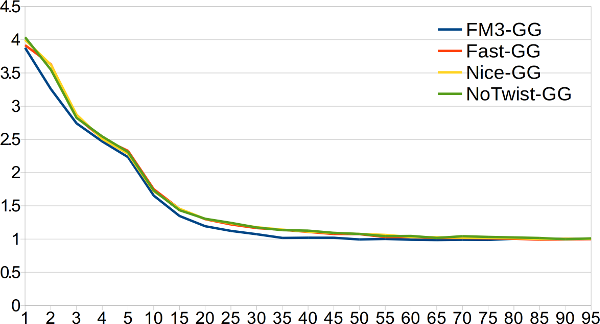}}
	\hspace{1pt}
	\subfloat[Black-hole graphs]{
		\includegraphics[width=0.48\linewidth]{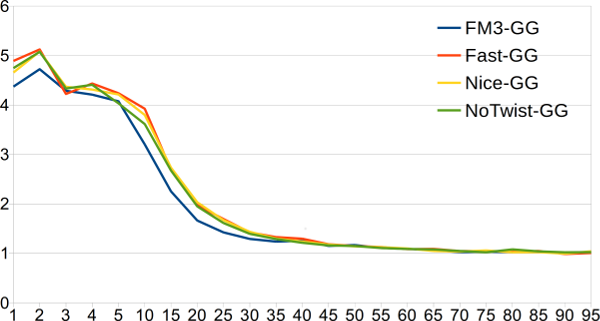}}	
	\caption{\label{fig:qualityratioLayouts}Comparison of average quality ratio of \DSS~over \RE~between FM$^3$, Fast, Nice and NoTwist layouts. The y-axis shows the average quality ratio $Q_{\DSS}/Q_{\RE}$.}
\end{figure*}

\end{enumerate}


\begin{figure*}[t]\centering
	\subfloat[Inverse time]{
		\includegraphics[width=0.48\linewidth]{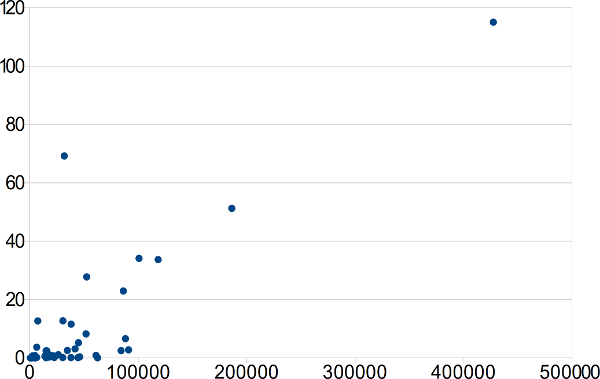}}
	\hspace{5pt}
	\subfloat[Resistance computation time]{
		\includegraphics[width=0.46\linewidth]{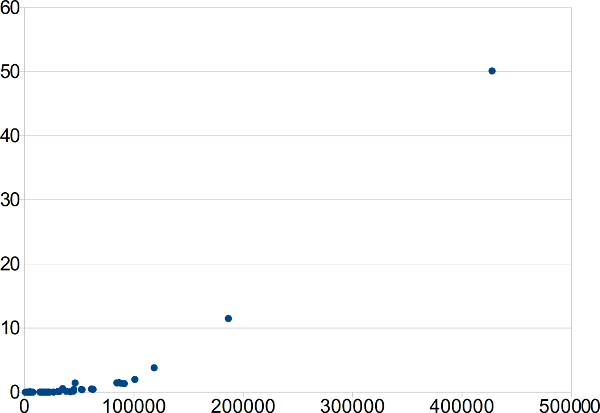}}	
	\caption{\label{fig:runtime}The running time of computing Moore-Penrose inverse (in minutes) and resistance values of all edges (in seconds).}
\end{figure*}

\subsection{Runtime}
\label{sec:runtime}

Although the main purpose of our investigation was to evaluate the \emph{effectiveness} of spectral sparsification,
some remarks about runtime are in order.

Fig.~\ref{fig:runtime}(a) illustrates runtimes. The x-axis shows the number of edges, and the y-axis shows the computation time in minutes. Fig.~\ref{fig:runtime}(b) shows the amount of time for (parallel) computing resistance values. The x-axis shows the number of edges, and the y-axis shows the computation time in seconds.

The dominant part of runtime is the computation of the Moore-Penrose inverse (and thus effective resistance); for this we used standard software~\cite{DBLP:conf/mm/HareSD11}.
For the defacto-benchmark graphs, computing the Moore-Penrose inverse takes 10.42 minutes on average. Graph \emph{can\_144} takes minimum time for the Moore-Penrose inverse calculation (0.0003 mins), and graph \emph{graph\_3} takes the longest time (115 mins).

%
%
%

\section{Concluding remarks \label{sec:conclusion}}

This paper describes the first empirical study of the application of spectral sparsification in graph visualization.

Our experiments suggest that spectral sparsification approaches (\DSS~and \SSS) are better than random edge approach.
Further, the results suggest some guidelines for using spectral sparsification:
\begin{itemize}
	\item \DSS~works better than \SSS~in practice.
	\item \DSS~and \SSS~give better quality metrics for grid-like graphs
          than for scale-free graphs.
	\item For sparsifications with low relative density (1\% to 20\%), \DSS~and \SSS~are considerably better than edge sampling. For relative density larger than 35\%, \RE~may be more practical, because it is simpler, faster, and produces similar results to \DSS~and \SSS.
\end{itemize}

\noindent
Future work includes the following:
\begin{itemize}
\item
Improve the runtime of these methods. For example,
Spielman and Srivastava~\cite{DBLP:journals/corr/abs-0803-0929} present a nearly-linear time algorithm that builds a data structure from which we can query the approximate effective resistance between any two vertices in a graph in $O(\log n)$ time. This would allow testing spectral sparsification for larger graphs.

\item
More extensive evaluation: our experiments compare spectral sparsification with random edge sampling, but not with the wide range of sampling strategies above. Further, extension to larger data sets would be desirable.

\item
In our experiments, quality is measured using an objective shape-based metric. It would be useful to measure quality subjectively as well, using graph visualization experts as subjects in an HCI-style experiment.

\end{itemize}

{\small \bibliography{paper,spectral}}
\bibliographystyle{splncs03}


\end{document}